\documentclass[aps,twocolumn,showpacs,superscriptaddress]{revtex4}

\usepackage{amsmath}
\usepackage{amssymb}
\usepackage{graphicx}
\usepackage{bm}

\newcommand{\be}{\begin{equation}}
\newcommand{\ee}{\end{equation}}
\newcommand{\ba}{\begin{eqnarray}}
\newcommand{\ea}{\end{eqnarray}}

\newcommand{\bfr}{{\bf r}}
\begin{document}
\title{Dipolar Poisson-Boltzmann Equation: Ions and Dipoles Close to Charged Surfaces}

\author{Ariel Abrashkin}
\affiliation{School of Physics and Astronomy, Raymond and Beverly
Sackler Faculty of Exact Sciences\\
Tel Aviv University, 69978 Ramat Aviv, Israel}

\author{David Andelman}
\affiliation{School of Physics and Astronomy, Raymond and Beverly
Sackler Faculty of Exact Sciences\\
Tel Aviv University, 69978 Ramat Aviv, Israel}

\author{Henri Orland}
\affiliation{Service de Physique Th\'eorique, CE-Saclay, 91191
Gif-sur-Yvette Cedex, France}


\begin{abstract}
We present an extension to the Poisson-Boltzmann model where the
dipolar features of solvent molecules are taken explicitly into
account. The formulation is derived at mean-field level and can
be extended to any order in a systematic expansion. It
is applied to a two-plate system with oppositely charged surfaces.
The ion distribution and profiles in the dipolar
order parameter are calculated and can result in a large correction
to the inter-plate pressure.

\end{abstract}

\pacs{61.20.Qg, 68.08.De, 82.60.Lf, 82.70.Dd}

\maketitle

Charged objects (ions, interfaces and particles) immersed in liquids
play a central role in electrochemistry, colloidal science and
biology ranging from electrolyte applications, stabilization of
colloidal suspensions, protein folding and its biological activity,
and even in protein
aggregation~\cite{DA1,RN1,israel,verwey,colloid_book,henderson,protein}.

The most commonly used model --- the Poisson-Boltzmann model (PB)
\cite{DA1,israel,verwey}
--- assumes point-like ions immersed in a continuum dielectric media
and treats the system in a mean-field approximation. The medium is
modeled by a homogeneous and isotropic dielectric constant. This
model is simple, elegant and efficient. It is in good agreement with
experiments for monovalent ions up to energies of order of $k_B T$.
However, careful measurements of the forces between two charged surfaces at
nanometric scale show strong deviation from the simple PB picture
\cite{israel}.
In particular, the assumption that the continuum dielectric medium
is homogeneous does not take into account
the strong dielectric response of water molecules around
charges. The discrete moments of water molecules will orient
themselves close to charged ions and surfaces giving rise to
hydration shells and to hydrophobic interactions, which can be
measured at short distances, for example, between two charged plates
(surface force balance apparatus). These hydration phenomena are very important
in many biological processes such as protein folding,
protein crystallization and interactions between charged biopolymers inside
the cell.

Most studies other than the PB rely on one of several theoretical
techniques. Monte Carlo (MC) \cite{MC1} or Molecular Dynamic
(MD) \cite{MD1} computer simulations take into account the discrete nature of the
dipolar molecules.
A second approach relies on liquid state theory, integral
equation and other methods \cite{liquid,podgornik}.
In simple planar geometry the latter gives good agreement with the
MC and MD simulations. However, all these methods are rather cumbersome
and involve heavy computation resources. In addition, they lack the
simple physical picture provided by a Poisson-Boltzmann type of
approach.

In this Letter we propose another approach called the  Dipolar
Poisson-Boltzmann (DPB). Unlike the PB model where the solution is
characterized  by a homogeneous dielectric constant, in the DPB model
we coarse grain the interaction of individual ions and dipoles
interacting together. This makes the DPB an analytic extension of
the PB formalism. Although it is done on a mean-field level, it
includes some aspects of the discrete nature of the dipolar solvent
molecules and how they modify the ion--solvent interactions. We
show that such corrections to the PB treatment are important in predicting dipolar
profiles close to charged surfaces and result in a strong
deviation from their average value.
Furthermore, the DPB model
can, in principle, be expanded to any desired higher order in a systematic expansion.

Consider a system composed of $N_d$ mobile dipoles each with a dipolar moment ${\bf p}$
and $I$ species of ions immersed in a continuum dielectric medium
with a weak dielectric response (the justification for this system set-up
is elaborated below),
$\epsilon\gtrsim \epsilon_0$,  $\epsilon_0$ being the vacuum
permittivity. Each ionic species has
$N_j$ ions of charge $q_j e$, $j=1\dots I$, where
$e$ is the electron charge. In
addition, the system includes a fixed charge distribution $\rho_f(\bfr)$.
The charge density
created by a point dipole ${\bf p}$ at point ${\bfr_0}$ is given
by $\rho_d(\bfr) = -{\bf  p} \cdot { \nabla} \delta(\bfr - {\bfr_0})$.
Thus, the total charge density is
\be \label{1} \rho(\bfr) = - \sum_{i=1}^{N_d} {\bf p}_i \cdot {\nabla}
\delta(\bfr - {\bfr_i})+ \sum_{j=1}^I \sum_{i=1}^{N_j} q_j e\delta (
\bfr -{\bf R}_i^{(j)}) +\rho_f (\bfr) \ee
where $\bfr_i$ denotes the position of  dipoles  of moment ${\bf  p}_i$ and
${\bf R}_i^{(j)}$ are the positions of  ions of type $j$.
The canonical partition function is given by
\ba
Z= \frac {1}{N_d! \Pi _{j=1}^I N_j!} \int \Pi_{i=1}^{N_d} d^3 \bfr_i d^3 {\bf p}_i \Pi_{j=1}^I
\Pi_{i=1}^{N_j} d^3 {\bf R}_i^{(j)} \nonumber \\
\times {\rm e}^{-\frac{\beta}{2}\int d^3\bfr d^3\bfr' \rho(\bfr) v_c
(\bfr-\bfr') \rho(\bfr')} \ea
where $v_c(\bfr)$ denotes the Coulomb potential.
Using a standard Hubbard--Stratanovich transformation,
\ba \label{3} Z&=& \int {\cal D}\phi(\bfr)\, \exp \left(
-\frac{\beta \epsilon}{2} \int d^3\bfr\ [\nabla
\phi(\bfr)]^2\right.
\nonumber\\
&+& \lambda_d \int d^3\bfr \,d^3 {\bf p} \ {\rm e}^{-i\beta {\bf
p}\cdot \nabla \phi} + \sum_{i=1}^{I}
\lambda_i \int d^3 \bfr \,{\rm e}^{-i \beta q_i e \phi}\nonumber \\
&& \left.- i \beta \int d^3 \bfr\ \phi(\bfr) \rho_f(\bfr) \right)
\ea
where $\epsilon=\epsilon_0\epsilon_r$ is the medium dielectric constant (in SI
units) and $\beta=1/T$ is the inverse temperature (where the
Boltzmann constant, $k_B$, is set to unity). The fugacities of the dipoles
and   $i^{th}$ ion species, $\lambda_d$ and $\lambda_i$, respectively,
are derived from the relations: $N_d = \lambda_d
\frac{\partial}{\partial \lambda_d} \log Z$ and $N_i = \lambda_i
\frac{\partial}{\partial \lambda_i} \log Z$.

Assuming that each molecular dipole has a fixed magnitude, $|{\bf p}|=p_0$
we sum now over the $\{{\bf p}\}$ degrees of freedom  and obtain
\ba
\label{4a}
Z&=& \int {\cal D}\phi(\bfr)\, \exp \left(-\frac{\beta \epsilon}{2} \int d^3\bfr\,
[\nabla \phi(\bfr)]^2 \right.\nonumber\\
&+&\lambda_d \int d^3\bfr\,
\frac{\sin{\beta p_0 |\nabla \phi}|} {\beta p_0 {| \nabla \phi}|}
+\sum_{i=1}^I \lambda_i \int d^3 \bfr\ {\rm e}^{-i \beta q_i e \phi}\nonumber\\
&& \left. -i \beta \int d^3 \bfr \,\phi(\bfr) \rho_f(\bfr) \right)
\ea

The DPB equation is obtained as the saddle-point of the action (\ref{4a})
[where we have used $\Psi(\bfr)= i \phi(\bfr)$
to denote the physical electrostatic potential]

\ba
\label{4}
-\epsilon \nabla^2 \Psi &=&
\sum_i \lambda_i q_i e\
 {\rm e}^{- \beta q_i e \Psi} + \rho_f(\bfr)
\nonumber\\
&+& \lambda_d p_0 {\nabla} \cdot \left[
\frac {{\nabla} \Psi}{|{\nabla} \Psi|}{\cal G}
\left(\beta p_0 |{\nabla} \Psi|\right) \right]
\ea
and the function ${\cal G}(u) =
 {\cosh u}/{u} -  {\sinh u}/ {u^2} $ is related to the
Langevin function ${\cal L}(u)=\coth u -1/u$ by ${\cal G}=(\sinh u/u) {\cal
L}$. One recognizes in (\ref{4}) the usual terms of the
Poisson-Boltzmann equation (the first two terms on the RHS), while
the last term is the divergence of the  polarization contributing to
the induced charge density.
The local polarization density (square bracket) in Eq.~(\ref{4}) is
the product of the dipole density, $\sinh u/u$, and the average dipole moment given (on a mean-field
level) by the Langevin function.

In the following we  study a dipolar solvent with 1:1 salt confined
between two {\em oppositely charged} planes \cite{antisymmetry}.
While the dipolar effects are small for
two similarly charged plates, they  are pronounced
for anti-symmetric plates and yield a
spatial variation of the dielectric constant. Choosing the charge density
to be $\mp \sigma$ for the two plates located at $z=\pm d/2$, the
potential, ionic profiles and  dipole density depend only on the $z$
coordinate  perpendicular to the planes and (\ref{4}) becomes
\ba \label{5} &-&\epsilon \Psi'' (z) =  -2 c_s e \sinh {\beta e
\Psi} +\sigma \delta(z+d/2) \nonumber\\ &-&\sigma \delta (z-d/2) ~+~
c_d p_0 \frac{d}{dz} \Bigl[{\cal G}(\beta p_0 \Psi') \Bigr] \ea
where we assume that the system is in contact with a reservoir containing
a dipolar fluid of concentration $c_d$ and salt  of
concentration $c_s$ so that $\lambda_d=c_d$ and $\lambda_s=c_s$.

The boundary condition at the $z=-d/2$ charged plane is
\be -\epsilon \Psi'_s =  c_d p_0 {\cal G}(\beta p_0 \Psi'_s)+
\sigma \label{bc_DPB}\ee
and  the electric field $E=-\Psi'$
is the same, for the anti-symmetric system, as on the other
plane. Note that the usual Neumann boundary conditions for the PB
equation includes now the polarization induced surface charges. We
find that for strong enough surface charge densities the induced
charge can be substantial and corresponds to a large modification of
the standard boundary condition.

From (\ref{5}) we obtain the first integral which is equivalent
to the contact theorem expression for the pressure difference
$\Pi=P_{\rm in}-P_{\rm out}$
\ba
\Pi=-\frac {\epsilon}{2} \Psi'^2(z)+ 2 c_s T (\cosh \beta e \Psi -1) \nonumber\\
-~c_d p_0 \Psi' {\cal G}(\beta p_0 \Psi') + c_d T \left(\frac {\sinh
\beta p_0 \Psi'}{\beta p_0 \Psi'} -1\right)\ea

This equation allows to express $\Psi(z)$ as a of function
$\Psi'$ and thus solves (\ref{5}) by a simple quadrature. The
first two terms in $\Pi$ are the usual PB  contributions,
the first being the electric field and the second  the
mixing entropy of the ions. The other two terms are the specific
terms of the DPB model. The first is the enthalpic contribution
related to the orientation of the dipoles in a local electric field.
The last term is the rotational entropy of the dipoles. The pressure
at any point $z$ is calculated with respect to the pressure exerted
by the bulk reservoir outside the plates.

Another way to interpret (\ref{5}) is to write it as a PB
equation with an effective field-dependent dielectric constant
$\epsilon^{\rm eff}(E)=\epsilon_0 \epsilon_r^{\rm eff}(E)$ replacing the $\epsilon$ on
the LHS.  The non-linear dielectric response is given by
\be\label{eps_eff}
 \epsilon^{\rm eff}(E)=\epsilon+ \frac{c_d p_0}{ E}{\cal G}(\beta
 p_0E)
 \ee
For weak fields one can expand the function ${\cal G}$ to first
order and obtain the standard PB equation
$\epsilon^{\rm eff} \Psi'' (z) \approx 2 c_s e \sinh{\beta e \Psi}$
with an effective homogeneous dielectric constant
\ba
\label{eps_linear}
\epsilon^{\rm eff} =\epsilon + \beta c_d {p_0^2}/{3 }
\ea

This result  for dielectric response of molecules with intrinsic
dipoles in dilute systems is well known \cite{jackson}. Since we are interested in
aqueous solutions, we have chosen as a fit parameter the molecular dipole moment of water to be
$p_0=4.86$~Debye (instead of the physical value
$p_0=1.85$). This allows us to obtain
$\epsilon_r^{\rm eff}=80$ for $\epsilon=\epsilon_0$ (vacuum permittivity)
and  $c_d=55$~M.
%

When the dipolar effects are strong (see below)  there is a crowding
of dipoles and ions between the plates, and their densities can  reach values higher
than close packing. To avoid this problem, we can generalize our
theory to take into account the finite molecular size \cite{itamar}.
Assuming that the 1:1 ions and dipoles are constrained on a lattice
of spacing $a$ (roughly equal to their molecular size), and imposing
the condition that each site of the lattice is occupied by only one
of the three species (incompressibility condition), the free energy
becomes
\ba
&-\beta F = \frac{\beta \epsilon}{2} \int d^3\bfr [\nabla \Psi(\bfr)]^2 \nonumber \\
&+\frac{1}{a^3}  \int d^3\bfr\, \log \left( c_d \frac {\sinh{\beta
p_0 |\nabla \Psi}|} {\beta p_0 |\nabla\Psi|}+ 2 c_s \cosh (\beta e
\Psi) \right) \ea
where $c_d+2c_s=a^{-3}$. Minimizing the above free energy,
the Modified Dipolar Poisson-Boltzmann (MDPB) equation is obtained
\ba
\label{9}
&-\epsilon\Psi''(z) =  \sigma \delta(z+d/2) - \sigma \delta (z-d/2) \nonumber \\
&+\frac {c_d p_0}{a^3}  \frac{d}{dz} \left[ \frac {{\cal L}(\beta
p_0 \Psi')}{{\cal D}}\right] - \frac {2 c_s e}{a^3} \frac{\sinh
{\beta e \Psi}}{\cal D} \ea where \be {\cal D} = {c_d
\frac{\sinh{\beta p_0 \Psi'}}{\beta p_0 \Psi'}+ 2 c_s \cosh (\beta e
\Psi)}
\ee
The presence of the denominator ${\cal D}$ in (\ref{9}) leads to
saturation of the local ionic and dipolar densities, which is quite important
close to charged boundaries. Without the dipolar effect
$p_0=0$, the MDPB equation reduces to the modified PB equation which
also displays an ionic saturation effect because of solvent entropy
\cite{itamar}.

A large deviation of the DPB treatment from the standard PB one may
occur in the strong E field regime. Such a case is presented now by
solving numerically eqs.~(\ref{5})--(\ref{bc_DPB})  for a system
composed of two planar surfaces located at $z=\pm d/2$, with
opposite surface charge densities $\mp \sigma$ and with small
amounts of 1:1 salt to avoid strong screening effects.
In this anti-symmetric system the potential at
the mid-plane vanishes, while the electric field there is non zero. The
DPB pressure, in turn, deviates substantially from its corresponding  PB
value due to the coupling between the dipole density and the non-zero
electric field. This is in contrast with a symmetric planar system
where the electric field vanishes at the mid-plane.

\begin{figure}[h!]
\begin{center}
\includegraphics[scale=0.49]{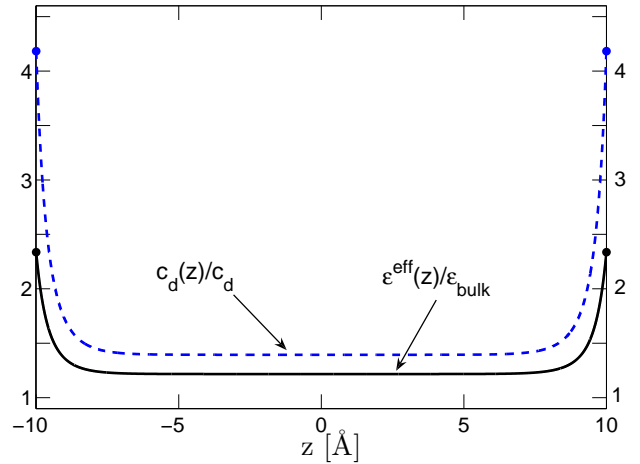}
\end{center}
\label{Fig1} \caption{The DPB rescaled dielectric constant $\epsilon^{\rm
eff}(z)/\epsilon_{\rm bulk}$ and the dipole density $c_d(z)/c_d$
profiles between two
oppositely charged plates at separation $d=20$\,\AA. The surface
charge density is $\sigma=\mp e/50$\,\AA$^2$. The reservoir contains
1:1 salt of concentration $c_s=10^{-5}$\,M and dipoles of density
$c_d=10$\,M. The dielectric constant is rescaled with respect to its
bulk value $\epsilon_{\rm bulk}=18.2$. The profiles have a strong
variation in the vicinity of the plates (up to 2\,\AA) and then
saturate to a value that is somewhat higher than their bulk values.}
\end{figure}

Had we modeled the water solvent as dipoles in vacuum ($\epsilon=\epsilon_0$), the  dipole density
in the mid region (see Fig.~1) would have reached unphysical values
above the close packing ones, because nothing in our model prevents over-crowding.
In order to avoid
this  artifact we use a background of low dielectric solvent
({\it e.g.}, $\epsilon_r=4.5$ for ether) and treat {\it explicitly} the
strong water dielectric response by the dipolar term in the DPB equation (\ref{5}).
In this fashion the water bulk density is lower than its close packing value, yielding a
dipole profile density which is higher than the bulk value but  below the close packing one.
Note that all other mixture enthalpic and entropic terms are
not considered at present~\cite{todo}.

In Fig.~1 we present the DPB profiles for the dipole density and local
dielectric constant between two charged plates with separation of
$d=20$\,\AA. The figure shows a strong accumulation of dipoles
between the charged plates leading to high effective dielectric constant. The profile of
the dipole density (dashed line) is rescaled by its bulk value. It
can be seen  that in the surface vicinity (up to about
2\AA), the density rises to above four times its bulk value due to
the strong attraction with the charged surface. In the mid-region
the density saturates at about 1.4 times its bulk value. The
corresponding local effective dielectric constant (solid line in the figure) can be
calculated from (\ref{eps_eff}). The
profile resembles that of the dipole density. In rescaled units, it
saturates at a value of about 1.2 in the mid-region and reaches about 2.3
at the surfaces.

Compared to a PB theory with the {\em same} bulk dielectric constant (\ref{eps_linear}),
which is
taken as constant throughout the system, the DPB demonstrates strong
deviations, not only in the surface proximity but also in its
saturated mid-range value (for strong enough $\sigma$ and/or small
$d$).

The ionic concentration is  much less affected by the presence of
the dipoles. We have computed the ion densities as a function of the
distance to the surface.  Because of the different boundary
condition, (\ref{bc_DPB}), the ionic density is strongly
suppressed at the surface with respect to PB (to about half of its
original value). However, it comes back to its PB value at distances as close
as 0.5 \AA\ from the surface.

In Fig.~2 we plot the relative osmotic pressure difference
$(\Pi_{\rm DBP}-\Pi_{\rm PB})/\Pi_{\rm PB}$ as a function of the
surface separation $d$. The pressure is a global quantity, and is
sensitive to the strength of the electric field throughout the
system rather than to its value on the surface. As a result, $\Pi_{\rm DPB}$ 
deviates strongly from $\Pi_{\rm PB}$
for small $d$, while $\Pi_{\rm DPB}\approx \Pi_{\rm PB}$  
at larger separation. 

\begin{figure}[h!]
\begin{center}
\includegraphics[scale=0.49]{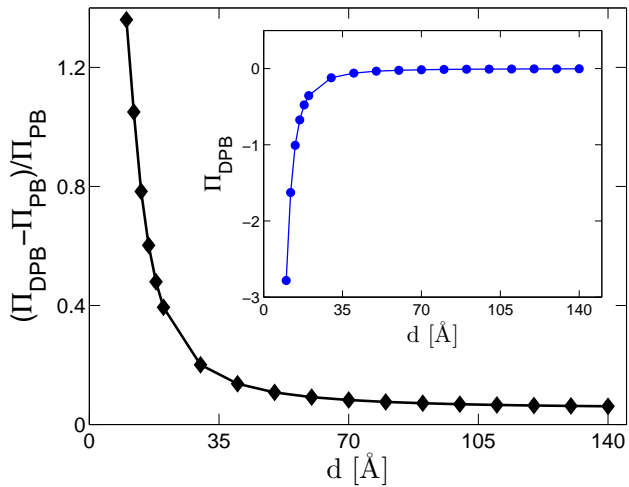}
\end{center}
\label{Fig2} \caption{The DPB calculated pressure $\Pi$ as function
of the inter-plate separation $10\le d \le 140\,{\rm \AA}$ for two oppositely
charged plates. All other system parameters are as in Fig.~1. We compare
the DPB with the usual PB models by plotting their relative
difference $(\Pi_{\rm DBP}-\Pi_{\rm PB})/\Pi_{\rm PB}$. In the inset
the DPB pressure is plotted in units of $10^{-2}\,k_B T/{\rm \AA}^3$.  }
\end{figure}

 We have presented  an analytical modification of the PB equation
by including the dipole degrees of freedom. We calculated the
correction to the potential, electric field and densities for a
system of two oppositely charged plates (Fig.~1). The results are
compared with those of the usual PB equation with an effective
dielectric constant. We find that when the electric field is strong
($p_0 E \approx k_B T$),  there are strong deviations from the PB
model. The spatial dependence of the dielectric constant
signals an ordering of the dipoles at the surfaces. This spatial
dependence is also a signature of non-linearity in the dielectric
response. The inter-plate pressure is sensitive to the value of the
electric field at the mid-plane and can deviate considerably from
the PB results for small enough separation and/or large surface
charges (Fig.~2).

The formalism presented here is general but was applied at a
mean-field level. The PB equation is analytically modified by the
dipole degrees of freedom. As a mean-field approximation it also has its
usual limitations. First, it lacks correlation effects. In addition,
it does not treat correctly the finite size of the ions and dipoles
and the densities of both can reach unphysical high values in high E
fields. The latter limitation can be remedied by including the hard
core of ions and dipoles, (\ref{9}), and will be published
elsewhere \cite{todo}.

The results presented in this paper can be verified experimentally
by using, for example, the Surface Force Balance (SFB) apparatus. Recent SFB experiments
\cite{klein} have been performed on asymmetrically charged surfaces.
The range of inter-surface separations that we used can be explored
using the SFB technique. What is needed, however, are careful
studies of mixtures of different dielectric solvents in order to
extract the dipole contribution to the osmotic
pressure. This systematic set of experiments may shed light on the
short-range hydrophobic effect and hydration forces.

{\em Acknowledgements.~~~~} We thank L. Arazi, D. Ben-Yaakov, Y. Burak, D. Harries
and S. Safran
for helpful discussions and comments.
Support from the Israel Science Foundation (ISF) under
grant no. 160/05 and the US-Israel Binational Foundation (BSF) under
grant no. 287/02 is gratefully acknowledged.




\end{document}